\input ./style/arxiv-general.cfg
\documentclass[aoas,MSNbibl,nameyear,seceqn,dvips]{arximspdf}
\makeatletter
   \@ifpackageloaded{graphicx}{}{\usepackage{graphicx}}
\makeatother
\doi{10.1214/15-AOAS868}% Updated by VTEXPTS2LaTeX.exe, 07.10.2015 15:45
\volume{9}
\issue{4}
\pubyear{2015}
\firstpage{2073}
\lastpage{2089}
\docsubty{FLA}


\begin{document}
\begin{frontmatter}

\title{Modeling competition between two pharmaceutical drugs using innovation diffusion models\thanksref{T1}}
\runtitle{Modeling competition between two pharmaceutical drugs\hspace*{10pt}}
\thankstext{T1}{Supported by Universit\`a degli Studi di Padova,
Progetto di Ricerca di Ateneo CPDA118082/11:
``Competition among Innovation Diffusion Processes.''}

\begin{aug}
% Corresponding author: Cinzia Mortarino - mortarino@stat.unipd.it% Updated by VTEXPTS2LaTeX.exe, 08.10.2015 13:22
%Updated by VTEXPTS2LaTeX.exe, 07.10.2015 15:45
\author[A]{\fnms{Renato} \snm{Guseo}\ead[label=e1]{renato.guseo@unipd.it}}
\and
\author[A]{\fnms{Cinzia} \snm{Mortarino}\corref{}\ead[label=e2]{mortarino@stat.unipd.it}}
\runauthor{R. Guseo and C. Mortarino}
\affiliation{University of Padova}
\address[A]{Department of Statistical Sciences\\
University of Padova\\
via Cesare Battisti, 241\\
35128 Padova\\
Italy\\
\printead{e1,e2}}
%\runauthor{}
%\dedicated{}
\end{aug}

% HISTORY:
%
\received{\smonth{3} \syear{2015}}% Updated by VTEXPTS2LaTeX.exe,
%07.10.2015 15:45
%
\revised{\smonth{7} \syear{2015}}% Updated by VTEXPTS2LaTeX.exe,
%07.10.2015 15:45

% ABSTRACT
%
\begin{abstract}
The study of competition among brands in a common category is
an interesting strategic issue for involved firms. Sales monitoring and
prediction of
competitors' performance represent relevant tools for management. In
the pharmaceutical
market, the diffusion of product \emph{knowledge} plays a special role,
different from the role it plays in other competing fields.
This latent feature naturally affects the evolution of drugs' performances
in terms of the number of packages sold. In this paper, we propose an
innovation diffusion model
that takes the spread of knowledge into account.
We are motivated by the need of modeling competition of two
antidiabetic drugs in the Italian market.
\end{abstract}

% KEYWORDS
\begin{keyword}
\kwd{Competition}
\kwd{innovation diffusion}
\kwd{dynamic market potential}
\kwd{communication network}
\kwd{nonlinear regression}
\end{keyword}
\end{frontmatter}

%s1 #&#
\section{Introduction} \label{sec:Intro}

The diffusion of an innovation often has to cope with the rise of many
competitors
that generate huge competitive effects, expansion or contraction in the market's
potential size, changes in the evolutionary dynamics of certain brands,
increases or
decreases
in life cycle length, and anticipation of the time of entry of additional
products in the market. These effects can be modeled only if they are
included in a
single complex system that can correctly identify competition and
contextual forces.

We cannot observe the complex system in which single agents (consumers)
may interact and share information regarding alternative technologies,
comparable solutions, similar devices, and so on. Instead, we observe the
resulting aggregate emergent dynamics (level reached by diffusion;
e.g., number of
packages sold),
and we base our analysis upon this level of observability.

Usually, the diffusion of products in a marketplace has a
limited time horizon defining particular finite life cycles with
different internal dynamics.
We observe poor performance at the beginning of the process after launch
due to limited acceptance of a newcomer to the market that interacts
with previous
knowledge and consumers' lifestyles. Similarly, but for different
reasons, we notice
a pronounced decrease in sales at the end of the commercial life cycle,
when the product is
perceived as an old, inefficient solution. Previous competing processes are
nonstationary and nonlinear due to chilling and saturating effects
within their
life spans.
Following this qualitative reasoning, for modeling and predictive
purposes, we may
exclude the direct use of ARMA-like (or VARMA) processes, which are strongly
based on weak stationary conditions after some differencing.

The pharmaceutical market is an important example of competition among
alternative drugs.
The products can differ to a great extent when they are based on
different active compounds,
or they can differ only at the commercial level when the same active
compound is sold
by competing firms. Moreover, this market differs from other markets,
since in many
countries the cost of essential/vital drugs is paid through a welfare
system. To some extent,
pricing does not directly influence
physicians' prescriptions. In addition, in Italy,
the Ministry of Health and pharmaceutical firms negotiate the price to
be paid by the national health service.

The aim of this paper is to build and apply a competition model for
pharmaceutical drugs (source: IMS Health Italy). In particular,
we focus on a pair of drugs with the same active
compound, based on glimepiride. This is a situation of substitute
products (brands)
competing for the same patients. The results will be compared with the
outcomes obtained by applying alternative models. In this case, we
emphasize that
an ``explicative'' model, in addition to describing the data and
providing reliable forecasts,
should highlight the key features of the competition among the analyzed drugs.
This is a further reason, beyond nonstationarity, not to rely on
traditional time series approaches.

A specific method for studying the dynamics of these special markets is
based on
two steps. First, to detect the mean trajectory of the processes, we use
the diffusion of innovation methodology,
which is strongly related to system analysis and epidemiological
modeling tools.
Second, to take into account seasonal autoregressive or moving average effects,
we perform an analysis of residuals, thereby improving short-term prediction.

The models due to Bass and colleagues
%(\citet{bass:69},
[\citeauthor{bass:69} (\citeyear{bass:69}),
%\citet{bass:94})
Bass, Krishnan and Jain (\citeyear{bass:94})]
represent an
essential step for the development of aggregate univariate diffusion
patterns, and a huge number of
extensions have originated from them
[see, among others,
%\citet{meade},
\citeauthor{meade}, (\citeyear{meade}),
%\citet{Peres2010}).
Peres, Muller and Mahajan (\citeyear{Peres2010})].

Conversely, the main contributions pertaining to competition modeling
are rather sparse.
%Krishnan et al,
\citet{krishnan:00},
%Savin and Terwiesch,
\citet{savin} and, recently, \citet{EJOR} and \citet{IMA} have
described competition with a differential representation
admitting a closed-form solution.
The differential representation is typical of the models proposed in
quantitative marketing literature,
where an aggregate parsimonious description of real adoption processes, based
on interpretable parameters, is essential
to capture relevant features, deduce related managerial implications,
and predict the future evolution of the market under study. The
simplicity of the model's structure is
obtained by introducing plausible assumptions regarding the behavior of
the agents
playing a role within the market.
In addition, a tractable solution for estimation and prediction
makes it easy to validate the model through aggregate sales data.
The relevant issue in this research topic is to build an adequately
large set of models to describe
the different characteristics of the diffusion process. Confirmation or
rejection of the assumptions
underlying
each model is then attained by fitting available observed data and
comparing the models' performances.

Models available in the literature to describe diffusion of competing
products in
a common market assume that the asymptotic market potential---that is,
the total
number of adoptions a product will ultimately reach at the end of its
life cycle---is
invariant throughout the life cycle from the products' launch.
However, this assumption is almost always unrealistic. In general,
knowledge and awareness
of a product are not immediately disseminated throughout eligible adopters
upon the entrance of a pioneering brand into the market. Moreover, new
brands are
often followed by competitors,
whose launch may affect awareness of the earlier products.
The topic arises from the consideration that awareness of a product
and adoption are themselves diffusion processes. Awareness is a latent
prerequisite for
adoption, and the degree of penetration of a product into the market is
limited by the
degree of diffusion of knowledge regarding its existence and
properties. For this
reason, the market potential would be better described as a dynamic process
than as a fixed constant, as discussed
in \citet{guseo:2009} for the univariate case without competition
effects.

In Section~\ref{sec:existingmodels} we briefly illustrate the standard
Bass model
[\citeauthor{bass:69} (\citeyear{bass:69})]
with its extension [\citeauthor{guseo:2009} (\citeyear{guseo:2009})],
that introduces a dynamic market potential.
The underlying reasons motivating this extension are also presented.
In Section~\ref{sec:newmodel} we discuss how the competition model
proposed in \citet{IMA}
can be extended to incorporate dynamic market potential. In Section~\ref
{sec:application}
we illustrate the application of the new model to the description of
competition between two antidiabetic drugs.
In Section~\ref{sec:comparison} we discuss the improvement obtained for these
data with the proposed dynamic potential model and
contrast it with the more common constant hypothesis and with
alternative dynamic structures.
In Section~\ref{sec:discussion} we present concluding remarks.

%s2 #&#
\section{A possible form for dynamic market potential} \label
{sec:existingmodels}

The simpler form of a univariate diffusion of an innovation model is
given by
the Bass model [\citeauthor{bass:69} (\citeyear{bass:69})]. The
differential representation
is defined through the following equation:
%
%e2.1 #&#
\begin{equation}
\label{BM} z^\prime(t)= m \biggl[p+q \frac{z(t)}{m} \biggr]
\biggl[1-\frac
{z(t)}{m} \biggr] = \biggl[p+q \frac{z(t)}{m} \biggr]
\bigl[m-{z(t)} \bigr],
\end{equation}
where $z(t)$ and
$z^\prime(t)=\partial z(t)/\partial t$ represent the mean cumulative
sales and the
mean instantaneous sales at time $t$, respectively.
Parameter $m$ is the fixed market potential
(the asymptotic level of cumulative sales or the total number of adoptions
at the end of the life cycle), that is, $m=\lim_{t\rightarrow\infty} z(t)$.

Equation (\ref{BM}) makes explicit that, at each time point, the
increase in instantaneous mean sales is proportional to the residual
market, $m-z(t)$.
The proportionality factor is affected by a fixed effect, $p$, and by a
time-varying effect, $qz(t)/m$.
The former, called the innovative coefficient, is independent of the
degree of diffusion
reached.
The higher the value of $p$, the more rapid the takeoff of the life
cycle, describing a
process in which exogenous factors, such as advertising or
institutional communication efforts,
push a product's diffusion.
The latter effect, $qz(t)/m$, depends upon the degree of saturation of
the market and
describes, through the interaction $z(t)[m-{z(t)}]$, how word-of-mouth
from previous
sales promotes further diffusion. The coefficient $q$ is called the
imitative coefficient.
The higher the value of $q$, the more important word-of-mouth is in
increasing diffusion. As $z(t)$ approaches $m$, the residual market,
$m-z(t)$, collapses and instantaneous mean sales, $z^\prime(t)$,
reduces to zero.

Under the initial condition $z(0)=0$, and defining $z(t)=0$ for $t<0$, the
explicit solution of equation (\ref{BM}) is
%
%e2.2 #&#
\begin{equation}
\label{BMsol} z(t)= m \frac{1-e^{-(p+q)t}}{1+ ({q}/{p}) e^{-(p+q)t}}
= m w(t;p,q),\qquad  t>0, p,q>0,
\end{equation}
where
%
%e2.3 #&#
\begin{equation}
\label{zsum} w(t;p,q)=\frac{1-e^{-(p+q)t}}{1+({q}/{p})e^{-(p+q)t}}.
\end{equation}

%It must be noted that the differential representation is used here
%instead of a
%discrete version, that is, a difference equation.
%The discrete representation converges to the continuous differential
%equation when the time interval tends to zero.
Continuous-time modeling is a common choice throughout the
diffusion of innovation literature, even when models are fitted to
weekly, monthly or quarterly
data.
This is partially because the involved variables are measured
continuously over time, even if, for administrative reasons, data are
recorded at discrete times.
In addition,
\citet{putsis}
conducts a detailed comparison and emphasizes that using seasonally
adjusted quarterly
data results in better estimates than using annual data.
In contrast, moving from quarterly to monthly data produces only
marginal statistical improvement.
\citet{Bosw} indicate that the values of $p$ and $q$ in their
discretized version of the Bass model
correspond to those of the continuous time model used here whenever equally
spaced data are available.

Although model (\ref{BM}) and its successive extensions proved to be
extremely valuable in describing
innovation diffusion processes, all are limited by the fact that market
potential, $m$, is a fixed constant, and hence cannot evolve over time.
This assumption conflicts with the common perception that knowledge
may be time
dependent. Some attempts have been proposed in the literature to
overcome this
limitation.
In some papers, the dynamic market potential is modeled as a function
of exogenous
observed variables [see, e.g.,
%\citet{Kim}
\citeauthor{Kim} (\citeyear{Kim}),
and the included references]. In other cases, it is assumed to be a
function of time only
[e.g.,
%\citet{Sharif},
\citeauthor{Sharif} (\citeyear{Sharif}),
%\citet{Centrone},
\citeauthor{Centrone} (\citeyear{Centrone}),
%\citet{Meyer}).
\citeauthor{Meyer} (\citeyear{Meyer})].

Here, we follow the approach proposed in \citet{guseo:2009}. In
principle, the market potential
can be any function $m(t)$ that defines an upper bound for cumulative
sales $z(t)$,
that is, $z(t) \leq m(t)$ for all $t$.
However, a parsimonious and intuitive method for specifying the form of $m(t)$
arises when we examine the communication network spreading information
about the
product in question.
The number of potential adopters of a product can be thought of, at
each time
point, as the size of the
aware agents' group. We describe awareness of the product as knowledge
transmitted
through a network
that describes the specific contacts among agents who eventually ``speak''
about the product.
This approach is linked to the literature on social networks, often
represented with random graph models where nodes denote
individual social actors (agents) and edges denote specific
relationships between two
%actors, \citet{hand:2010}.
actors [\citeauthor{hand:2010} (\citeyear{hand:2010})].
Many contributions to the literature assume observability of the edges, either
complete or partial, through sample data.

In our approach, conversely, the communication network evolving over
time is latent and
does not have to be observed or described in detail; this is also due
to the high costs of reliable relational data collection. The focus is
instead on the number of informed agents (active nodes).
This is a key aspect, since we want to deal with all the situations
where the communication network is \emph{product-specific}
(people usually choose to talk with someone---and not with someone
else---according to the topic of the conversation).
In these situations, the content-driven network is totally
unobservable, or it is
very difficult to obtain reliable \emph{pertinent} data.

The formalized structure of such a network is described in \citet
{guseo:2009}, where
the authors explain in detail how
this interpretation may lead to the following dynamic market potential function:
%
%e2.4 #&#
\begin{eqnarray}
\label{mdit} m(t)&=&K \sqrt{\frac{1-e^{-(p_c+q_c)t}}{1+ ({q_c}/{p_c})
e^{-(p_c+q_c)t}}}
\nonumber
\\[-8pt]
\\[-8pt]
\nonumber
& =& K \sqrt{w(t;p_c,q_c)},\qquad
K, p_c, q_c>0, t>0,
\end{eqnarray}
where $K$ is the upper asymptotic potential (directly related to the network's
size), $K=\lim_{t\rightarrow\infty} z(t)$, and
$p_c$ and $q_c$ are evolutionary
parameters describing how fast communication spreads {through} the
network. In particular,
for large values of $p_c$ and $q_c$, the dynamic market potential
$m(t)$ rapidly approaches $K$.

The expression under the square root in equation (\ref{mdit}) represents
the core of the Bass model
%(\citet{bass:69})
[\citeauthor{bass:69} (\citeyear{bass:69})]
% (Equation (\ref{BMsol}))
describing the latent diffusion process of communication. This
is an S-shaped curve, a distribution function, whose peakedness
varies according to the product's communication features.

The model proposed by \citet{guseo:2009} extends
the Bass model (\ref{BMsol}) in the following manner:
%
%e2.5 #&#
\begin{eqnarray}
\label{coev} z^\prime(t) = m(t) \biggl[p_s+
q_s \frac{z(t)}{m(t)} \biggr] \biggl[1-\frac
{z(t)}{m(t)} \biggr]+
z(t)\frac{m^\prime(t)}{m(t)},
\nonumber
\\[-8pt]
\\[-8pt]
 \eqntext{p_s, q_s>0, t\geq0,}
\end{eqnarray}
where $z(t)$
represents the mean cumulative sales, as in equation (\ref{BM}), and
$m(t) \geq z(t)$
may be defined as in (\ref{mdit}). The new parameters $p_s$ and $q_s$
are evolutionary
parameters that describe how fast the product is adopted (whereas, as
mentioned above,
$p_c$ and $q_c$ are related to knowledge spread).

The final term of equation (\ref{coev}) requires close examination to
understand its meaning.
This component enables us to take into account a self-reinforcing effect
that is common within marketing behavioral studies
[see, e.g., \citeauthor{sydow} (\citeyear{sydow}), a recent contribution on
self-reinforcing processes]. The standard adoption process
described in the first part of the equation is enhanced whenever the
market is growing faster.
In other words, an acceleration of the number of informed people [the
network's size, $m(t)$] further induces people to adopt. More
generally, excluding
assumption (\ref{mdit}), $m^\prime(t)$ may be negative when $m(t)$ is
nonmonotonic,
thereby introducing a shrinking effect on instantaneous sales due to a
decreasing market potential.

%s3 #&#
\section{The proposed model} \label{sec:newmodel}

The proposed model describes the diffusion of two competing brands.
They are supposed to be sufficiently similar to share a common market potential,
whose size grows in time as described in
Section~\ref{sec:existingmodels}. The assumption of a common market
potential is suitable in situations
where the products are substitutes competing for the same
adopters.
Whenever competition concerns products that are sufficiently different
to preserve product-specific market potentials, the Lotka--Volterra models
should be preferred, although these structures do not allow a
closed-form solution
[\citeauthor{abramsoms:98} (\citeyear{abramsoms:98})].

We denote the mean cumulative sales at time $t$ of brand $i$ by
$z_i(t)$, $i=1,2$,
and the
instantaneous mean sales by $z^\prime_i(t)=\partial z_i(t)/\partial t$,
$i=1,2$. We
now describe the category sales, $z(t)=z_1(t)+z_2(t)$, by separately
describing the two brands
constituting the
category. The model is given by
%
%e3.1 #&#
\begin{eqnarray}\label{1}\quad
z_1^\prime(t) &=& m(t) \biggl[p_1+(q_1+
\delta) \frac{z_1(t)}{m(t)} +q_1\frac{z_2(t)}{m(t)} \biggr] \biggl[1-
\frac{z(t)}{m(t)} \biggr]+ z_1(t)\frac{m^\prime(t)}{m(t)},
\nonumber
\\[-8pt]
\\[-8pt]
\nonumber
z_2^\prime(t) &=& m(t) \biggl[p_2+(q_2-
\delta) \frac{z_1(t)}{m(t)} +q_2\frac{z_2(t)}{m(t)} \biggr] \biggl[1-
\frac{z(t)}{m(t)} \biggr]+ z_2(t)\frac{m^\prime(t)}{m(t)},
\end{eqnarray}
where $z(t) \leq m(t)$, for all $t$.

To obtain an equivalent formulation of model (\ref{1}), that may be
more comparable
with the univariate Bass model, we can
rearrange the terms in the following manner:
\begin{eqnarray*}
z_1^\prime(t) &=& m(t) \biggl[p_1+
q_1 \frac{z(t)}{m(t)} +\delta\frac
{z_1(t)}{m(t)} \biggr] \biggl[1-
\frac{z(t)}{m(t)} \biggr]+ z_1(t)\frac
{m^\prime(t)}{m(t)},
\\
z_2^\prime(t) &=& m(t) \biggl[p_2+(q_2-
\delta) \frac{z(t)}{m(t)} + \delta\frac{z_2(t)}{m(t)} \biggr] \biggl[1-
\frac{z(t)}{m(t)} \biggr]+ z_2(t)\frac{m^\prime(t)}{m(t)}.
\end{eqnarray*}

In equation (\ref{1}), we may observe innovators' effects
(parameters $p_1$ and $p_2$)
%(parameters $p_1>0$ and $p_2>0$)
and word-of-mouth effects (parameters $q_1$, $q_2$ and $\delta$). These
parameters may be different for the two competitors to describe
products
with different strengths in the market.
Observe that this structure is similar to the model used in \citet{IMA},
allowing for
within-brand word-of-mouth ($q_1+\delta$ and $q_2$ for the two brands,
resp.)
that may be different from cross-brand word-of-mouth ($q_1$ and
$q_2-\delta$).
{In} other words, this model is able to deal with situations in which
word-of-mouth functions
asymmetrically for the two products.
In \citet{IMA}, however, unlike the proposed model,
$m(t)$ was supposed to be constant
throughout the life cycle: $m(t)=m$ for all $t$.

The final additive terms in equation (\ref{1})---which would obviously vanish
for a constant $m(t)$---represent a \emph{self-reinforcing} component,
as described in
the previous section.
The mean sales of both products are accelerated when $m(t)$
grows faster, that is, when awareness of the product category spreads
rapidly based on
the
collective behavior of agents.
Conversely, the mean sales are further
reduced by a shrinking potential induced by unfavorable
signals.
In the latter case, $m(t)$ could also be a
nonmonotonic function, and the
self-reinforcing term could be negative when the market
potential undergoes a contraction.

Notice that the sum of the equations in (\ref{1}) is equal to model
(\ref{coev}). Moreover, this model can also be used with an expression for
$m(t)$ that is
different from equation (\ref{mdit}).

Let us define $p_s=p_1+p_2$ and $q_s=q_1+q_2$. Through $w(t;p_s,q_s)$,
defined in~(\ref{zsum}),
%$$
%w(t;p_s,q_s)=\frac{1-e^{-(p_s+q_s)t}}{1+
%\frac{q_s}{p_s}e^{-(p_s+q_s)t}}, p_s, q_s>0,
%$$
and
%
%e3.2 #&#
\begin{equation}
y(t)= 1+\frac{q_s}{p_s}w(t;p_s,q_s) =
\frac{1+{q_s}/{p_s}}{1+
({q_s}/{p_s})e^{-(p_s+q_s)t}}, \label{newpost}
\end{equation}
it is proven in Appendix 1 [\citeauthor{suppA} (\citeyear{suppA})] that,
for any $m(t)$, the closed-form solution of the system
(\ref{1}) is
\begin{eqnarray}\qquad
\label{z1diasol} z_1(t) & = & m(t) \biggl\{ \frac{{q_{1}}}{q_s-\delta}
w(t;p_s,q_s) + \biggl[\frac{p_s}{\delta} \biggl(
\frac{p_{1}}{p_s}- \frac
{q_{1}}{q_s-\delta} \biggr) \biggr] \bigl[y(t)^{\delta/q_s} -1
\bigr] \biggr\},
\nonumber
\\[-8pt]
\\[-8pt]
\nonumber
z_2(t) & \stackrel{} {=} & m(t) \biggl\{ \biggl( \frac{{q_{2}-\delta}}{ q_s-\delta}
\biggr)w(t;p_s,q_s) + \biggl[\frac{p_s}{\delta} \biggl(
\frac{p_{2}}{p_s}- \frac{q_{2}-\delta
}{q_s-\delta} \biggr) \biggr] \bigl[y(t)^{\delta/q_s} -1
\bigr] \biggr\},
%\label{z2diasol}
\end{eqnarray}
when $\delta\neq0$ and $\delta\neq q_s$.
When $\delta=q_s$, the solution reduces to
%
%e3.3 #&#
\begin{eqnarray}\label{z1diasoldq}
z_{1}(t) &=& m(t) \biggl[ \biggl(\frac{p_{1}}{p_s}-
\frac{q_{1}}{q_s} \biggr) w(t;p_s,q_s) +
\frac{q_{1}p_s}{q_s^2}y(t)\ln y(t) \biggr],
\nonumber
\\[-8pt]
\\[-8pt]
\nonumber
z_2(t) &=& m(t) \biggl[ \biggl(1-\frac{p_{1}}{p_s}+
\frac{q_{1}}{q_s} \biggr) w(t;p_s,q_s) -
\frac{q_{1}p_s}{q_s^2}y(t) \ln y(t) \biggr],
\end{eqnarray}
while in the special case $\delta=0$, we obtain
%
%e3.4 #&#
\begin{eqnarray}\label{z1diasold0}
z_1(t) &=& m(t) \biggl[ \frac{ q_{1}}{ q_s} w(t;p_s,q_s)
+ \frac{p_s}{q_s} \biggl( \frac{p_{1}}{p_s}- \frac{q_{1}}{q_s} \biggr) \ln
y(t) \biggr],
\nonumber
\\[-8pt]
\\[-8pt]
\nonumber
z_2(t) &=& m(t) \biggl[ \frac{ q_{2}}{ q_s} w(t;p_s,q_s)
+\frac{p_s}{q_s} \biggl( \frac{p_{2}}{p_s}- \frac{q_{2}}{q_s} \biggr) \ln
y(t) \biggr].
\end{eqnarray}

The solutions for the mean cumulative sales enable us to use a
nonlinear regression model with dependent variables
given by the \emph{observed} cumulative sales of the two
brands.\setcounter{footnote}{1}\footnote{An alternative approach
using instantaneous sales is described in Appendix 2 [\citeauthor
{suppA} (\citeyear{suppA})].}
A reasonable and robust inferential methodology for estimating and testing
the performance of this structure may be
implemented through the regression model
%
%e3.5 #&#
\begin{equation}
\label{modregr} v_i(t) = z_i(t) + \varepsilon_i(t),\qquad
i=1,2,
\end{equation}
where $v_i(t)$ represents the observed cumulative sales data for each of
the two products
and $z_i(t;\beta)$
denotes the mean cumulative
functions (\ref{z1diasol})
depending on the vector of parameters
$\beta=\{K, p_c, q_c, p_1, q_1, p_2, q_2, \delta\}$ and on time $t$.
Henceforth, we use either the notation $z_i(t)$
or $z_i(t;\beta)$
to make explicit the dependence of the functions (\ref{z1diasol}) upon
$\beta$ parameters. Here, we assume that $m(t)$ is modeled
as in~(\ref{mdit}). In the rest of the paper, we will denote model (\ref
{modregr})---with
$m(t)$ specified as in (\ref{mdit})---with the expression Competition
Dynamic Market Potential (CDMP) model.
The residual term $\varepsilon_i(t)$ is usually a white noise or a
more complex stationary process if
seasonality or autoregressive aspects are included as stochastic components.
The joint estimate of $\beta$ is obtained with a single model where
$v_1(t)$ and $v_2(t)$ are stacked. This estimate could be generated
using the Beauchamp and Cornell technique
%\citet{beaucorn}.
[\citeauthor{beaucorn} (\citeyear{beaucorn})].
However, recent results show that it is advisable
to use ordinary nonlinear least squares
[\citeauthor{COST} (\citeyear{COST})].
Note that estimation through nonlinear
least squares does not require assumptions regarding the distribution of
$\varepsilon_i(t)$.
%\citet{COST}.
The nonlinear predicted values describe the mean trajectories of the
competing processes, that is, $z_1(t)$ and $z_2(t)$.

We propose a detailed simulation study in Appendix 5 [Guseo and Mortarino 
(\citeyear{suppA})]
to assess the performance of the CDMP model under different values of
the noise-to-signal
ratio when the latent market potential is correctly specified. We also consider
a further improvement
in the analysis of the robustness of the CDMP model for alternative
$m(t)$ structures.

A different approach, based on a stochastic version of equation (\ref{1})
including an error term with suitable assumptions, may be extremely
complex. This approach
is tractable, to our knowledge, only for simpler models such as the Bass
model [\citeauthor{Bosw} (\citeyear{Bosw})]. However, as mentioned in
Section~\ref{sec:existingmodels},
the Bass model is too simple a structure to describe complex markets.
\citet{Jha:2011} propose a stochastic differential equation model to
describe the adoption of newer successive
technologies. However, their work does not present a comparison with existing
deterministic models.
The comparison is essential to evaluate the effective gain of the
stochastic approach,
whose results are obtained through nonnegligible
assumptions regarding the stochastic component of the model, which may
be inappropriate for real
(not simulated) data sets.

%s4 #&#
\section{Antidiabetic drug sales case study} \label{sec:application}

\begin{figure}

\includegraphics{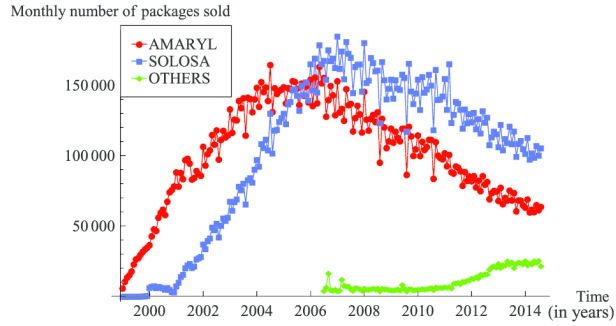}

\caption{Monthly sales data for Amaryl 2~mg and Solosa 2~mg. The series
of the sum of all the sales
of alternative products is also presented (source: IMS Health Italy).}
\label{fig:osservist}
\end{figure}

Amaryl (Sanofi--Aventis) and\break Solosa (Lab. Guidotti) are two
glimepiride-based drugs
used by people with type 2 diabetes.
Glimepiride belongs to the class of drugs known as sulfonylureas.
It lowers hyperglycemia by causing the body to release its natural insulin.
These drugs, at a dose of 2 mg, were launched in the Italian market in January
1999 and were for many years duopolists in the glimepiride market.
Figure~\ref{fig:osservist} shows monthly sales data (available until
{August 2014, for
a total of 188 data points}) for the
two drugs separately.
In addition, the figure depicts the series of the sum of all the
sales of alternative products (12 generic drugs) commercialized since 2006.
The more recent products have never represented an actual threat to the
two oldest brands.

These two drugs are perfect substitutes from the medical viewpoint, and
thus a
model with a common market potential appears to be an adequate
solution. Moreover,
in 1999, glimepiride represented a radical novelty in the Italian
market, since it
was the first type of sulfonylurea available.
Other dosages of the same drugs were launched much later, in 2006.
These considerations suggest that awareness of the properties and
efficacy of these drugs perhaps was not widespread among Italian
physicians in 1999. A dynamic market
potential seems conceivable for these data. The
complete impossibility of observing the communication network that spread
knowledge about glimepiride beginning in 1999 finally suggests that
the Guseo--Guidolin model (\ref{mdit}) could be an appropriate
tentative solution. Of course, only good agreement between the
available data and
functions
(\ref{z1diasol}), which incorporate these features, could confirm or
lead to
rejection of these assumptions.

%f1 #&#

Joint nonlinear regression of the two main competitors' cumulative
sales on
functions (\ref{z1diasol})---that is, the CDMP model, (\ref
{modregr})---gives rise to the
parameter estimates shown in Table~\ref{tab:par1}.

%t1 #&#
\begin{table}[b]
\caption{Estimation results for the CDMP model, (\protect\ref{modregr})}\label{tab:par1}
\begin{tabular*}{\textwidth}{@{\extracolsep{\fill}}lccc@{}}
\hline
& \textbf{Estimate} & \textbf{Standard error} &\multicolumn{1}{c@{}}{\textbf{95\% confidence interval}}\\
\hline
$K$ & $4.8669*10^7 $ & $2.5771*10^5 $\phantom{$-$} & ($4.81621*10^7$,
$4.9176*10^7$) \\
$p_{c}$ & \phantom{$-$}$2.3837*10^{-3}$ & $6.6814*10^{-5}$ & ($2.2523*10^{-3}$,
$2.5151*10^{-3}$) \\
$q_{c}$ & \phantom{$-$}$4.5235*10^{-2}$ & $4.6993*10^{-4}$ & ($4.4311*10^{-2}$,
$4.6159*10^{-2}$) \\
$p_1$ & \phantom{$-$}$3.2004*10^{-3}$ & $6.5762*10^{-5}$ & ($3.0711*10^{-3}$,
$3.3297*10^{-3}$) \\
$q_1$ & \phantom{$-$}$1.4277*10^{-2}$ & $3.2663*10^{-4}$ & ($1.3635*10^{-2}$,
$1.4920*10^{-2}$) \\
$p_2$ & $-7.9208*10^{-4}$ & $3.6160*10^{-5}$ & ($-8.6318*10^{-4}$,
$-7.2097*10^{-4}$) \\
$q_2$ & \phantom{$-$}$1.2709*10^{-3}$ & $5.5915*10^{-4}$ & ($1.7135*10^{-4}$,
$2.3704*10^{-3}$) \\
$\delta$ & $-2.2248*10^{-2}$ & $9.6448*10^{-4}$ & ($2.4145*10^{-2}$, $-2.0351*10^{-2}$) \\[6pt]
\multicolumn{4}{@{}l}{$R^2=0.99996$} \\
\hline
\end{tabular*}
\end{table}
The huge value of {$R^2 = 0.99996$} is unsurprising, given that we are
working with cumulative data and any S-shaped
fitting produces high determination indexes. A~standard approach advises
the use of the
$R^2$ measure only for comparative purposes, as will be described at
the beginning of Section~\ref{sec:comparison}.
In addition, the evaluation of the squared linear correlation
coefficient between observed
\emph{instantaneous} sales and fitted \emph{instantaneous} sales
yields a value of {0.9673}, which is extremely high.

\begin{figure}[t]

\includegraphics{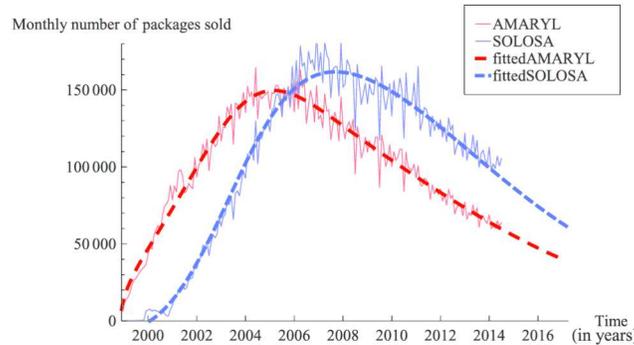}

\caption{Comparison of the monthly number of packages sold and fitted
values of
instantaneous sales using CDMP model, (\protect\ref{modregr}).} \label
{fig:fitted}
\end{figure}

\begin{figure}[b]

\includegraphics{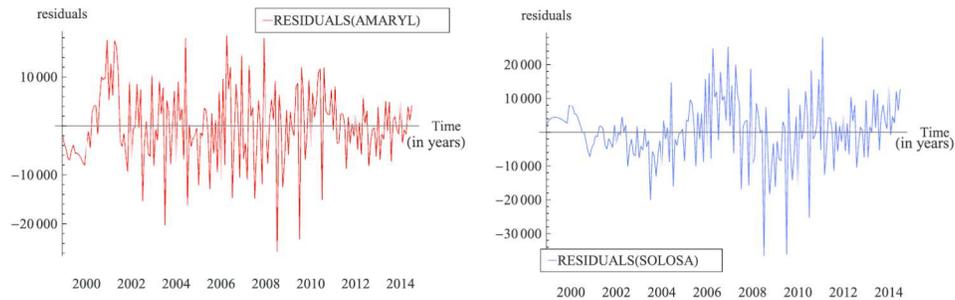}

\caption{Residuals for the two products (instantaneous sales scale).}
\label{fig:res}
\end{figure}

The agreement between the observed and fitted values can also be
assessed by examining
Figure~\ref{fig:fitted}. The two estimated profiles follow the
observations very well,
and discrepancies
(essentially due to seasonal effects) could easily be modeled using a SARMAX
approach characterizing the second step refinement for short-term prediction
[see Appendix 4, \citeauthor{suppA} (\citeyear{suppA})]. The analysis of
residuals is depicted in Figure~\ref{fig:res}.

%f2 #&#

%f3 #&#

\begin{figure}[t]

\includegraphics{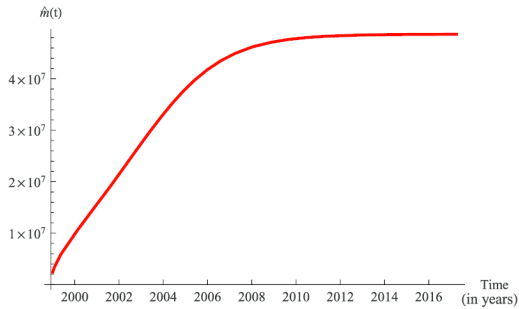}

\caption{Plot of the estimated market potential function, $\hat{m}(t)$.}
\label{fig:plotmt}
\end{figure}

Because we deal with consumables (i.e., repeatedly purchased goods),
$\hat{K}$ (49 million) represents an estimate of the total number of
packages of the two drugs that could be sold.
% and not the total number of potential adopters.
Figure~\ref{fig:plotmt}
depicts the estimated evolution of the common dynamic market potential, $m(t)$.
It is very far from a fixed $m$ {pattern}, since
knowledge of these drugs seems to have spread slowly among physicians.
This could be explained by the observation that a new active compound
(as glimepiride was in the Italian market in 1999) is accepted with
caution until side
effects are entirely disclosed.

%f4 #&#

If we focus on innovation parameters,
it is evident that this component did not play a significant role for Solosa,
and this may explain its slow start.
Lab. Guidotti, which launched Solosa, is a big
Italian company; however, its promotional strength could not compete
with the promotional efforts exerted by the international company
Sanofi--Aventis, which promoted Amaryl.

Imitative parameters have to be interpreted with reference
to the proposed model.
If we substitute the estimates in model (\ref{1}),
we obtain the following equations:
\begin{eqnarray*}
z_1^\prime(t) - z_1(t)\frac{m^\prime(t)}{m(t)} &
\propto& 0.0032 -0.0080 \frac{z_1(t)}{m(t)} + 0.0143 \frac{z_2(t)}{m(t)},
\\
z_2^\prime(t) - z_2(t)\frac{m^\prime(t)}{m(t)} &
\propto& -0.0008 +0.0235 \frac{z_1(t)}{m(t)} + 0.0013 \frac{z_2(t)}{m(t)}.
\end{eqnarray*}
Amaryl was sustained by a stronger innovation effect, and its cycle began
much more rapidly than its competitor's cycle (0.0032 vs. $-0.0008$).
Sanofi--Aventis is a much larger company than Lab. Guidotti, and the
former's promotional strength enabled an impressive start
to Amaryl's sales. However, Amaryl experienced a negative within-brand
word-of-mouth effect, in contrast with Solosa's positive effect
($-0.0080$ vs. 0.0013).
Both products were sustained by a positive cross-brand
word-of-mouth effect from the competitor, but the effect of this was
to increase Solosa's sales more strongly
(0.0235 vs. 0.0143).
This ultimately led Solosa to outsell Amaryl. Both drugs now appear to
be in a
declining phase of their life cycle, due to the appearance of other
active compounds
in the type 2 diabetes market.

%f5 #&#
\begin{figure}

\includegraphics{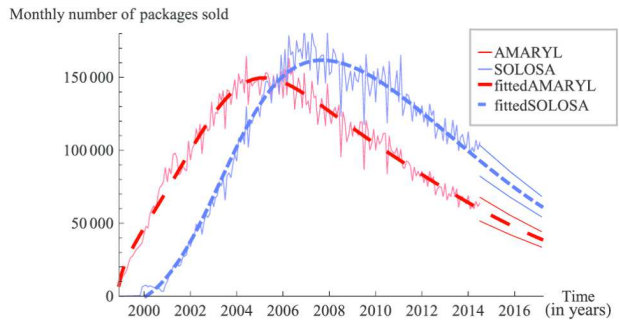}

\caption{Mean sales forecast and confidence bands for estimates based
on the CDMP model, (\protect\ref{modregr}).}
\label{fig:bandecum}
\end{figure}

Figure~\ref{fig:bandecum} illustrates predictive confidence bands for
the future
sales of the two products. Details regarding their construction are
given in
Appendix 3 [\citeauthor{suppA} (\citeyear{suppA})].

%s5 #&#
\section{Comparison with alternative models} \label{sec:comparison}

The efficacy of the proposed model in this application must be proven
with reference to alternative
models. As mentioned in the \hyperref[sec:Intro]{Introduction}, we will examine a set of
models to identify
which one performs better with available observations. The first alternative
to be considered is
a simpler model with constant market potential.
As mentioned
above, it is plausible that knowledge of the properties of the new
active compound
did not arise immediately at the products' launch. However, this
hypothesis should
be tested by examining whether a model with dynamic market potential,
$m(t)$, really improves the fitting.

The model proposed in \citet{IMA} fits this purpose since it can be
obtained by
(\ref{z1diasol}) with the only restriction $m(t)=m$.
All other features related to the evolution of the process
are the same for the two models.
Thus, we can claim that if model (\ref{modregr}) shows a
significantly better performance than \citeauthor{IMA}'s model
(\citeyear{IMA}), this proves
that the market potential for this category evolved
in a manner that differs significantly from the constant path. Note,
too, that
other models [e.g., those by
%\citet{krishnan:00}, \citet{savin},} \\ \added{\citet{libai:09},
%\citet{EJOR})}
\citeauthor{krishnan:00} (\citeyear{krishnan:00}),
\citeauthor{savin} (\citeyear{savin}),
\citeauthor{libai:09} (\citeyear{libai:09}), \citeauthor{EJOR} (\citeyear{EJOR})]
are nested within the \citet{IMA} model.
The $R^2$ value for the \citet{IMA} model equals 0.9988.
Since this model is nested within model (\ref{modregr}),
an $F$ test can be used to detect whether the gain from the
simpler model to the more complex model is significant.
As the first step,
the squared multiple partial correlation coefficient
%
%e5.1 #&#
\begin{equation}
\label{r2parz} \widetilde{R}^2 = \bigl(R^2_{M1}
- R^2_{M2}\bigr)/\bigl(1- R^2_{M2}
\bigr)
\end{equation}
is calculated (here, $R^2_{M2}$ denotes the determination index
of the reduced model that has to be compared to model $M1$).
A possible test to verify the significance of the $s$ parameters
of the $M1$ model that are not included in model $M2$ may be given
by
%
%e5.2 #&#
\begin{equation}
\label{fparz} % F = \frac{\widetilde{R}^2 (N-k)}{(1-\widetilde{R}^2) s},
F = \bigl[\widetilde{R}^2 (N-k)\bigr]/
\bigl[\bigl(1-\widetilde{R}^2\bigr) s\bigr],
\end{equation}
where $N$ denotes the number of observations used to fit the models
and $k$ is the number of parameters included in model $M1$.
%
%f6 #&#
\begin{figure}

\includegraphics{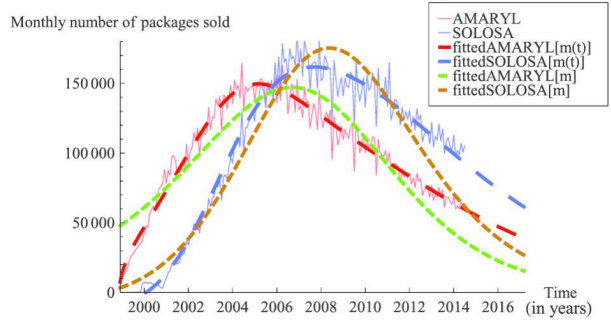}

\caption{Comparison of the fitted values for the monthly number of
packages sold using
the CDMP model, (\protect\ref{modregr}), and the model used in \citeauthor{IMA} \textup{(\citeyear{IMA}}).}
\label{fig:CFR}
\end{figure}
Under the null hypothesis of equivalence between models $M1$ and $M2$,
(\ref{fparz}) is distributed as a Snedecor's F with $(s, N-k)$ degrees
of freedom,
if the stochastic component of the regression model is normal i.i.d.
This may not be true for our case. Nevertheless, the F ratio (\ref
{fparz}) can be used as an approximate
robust criterion for comparing model $M2$ nested in $M1$,
by considering the well-known common robust threshold 4.
%\citet{guseo:07}.
%(\citeauthor{guseo:07}, \citeyear{guseo:07}).
Here, the test comparing model (\ref{1}) with \citeauthor{IMA}'s
(\citeyear{IMA}) model
assigns a huge value of $F = $ 5474.78 ($\widetilde{R}^2 =$ 0.9675),
demonstrating the relevance
of the extended (\ref{modregr}) model.

In Figure~\ref{fig:CFR}, the fitted values of model (\ref{modregr}) and
\citeauthor{IMA}'s (\citeyear{IMA}) model are compared. The rigidity of
a fixed market
potential makes the latter model
inadequate to describe these data; even worse, for larger
$t$ values, it shows a heavy underestimation that makes forecasts
totally unreliable.

Both the result of the $F$ test and the graphical comparison prove that
a constant market potential is not adequate to describe this market.
Given that conclusion, it could be interesting to see whether
alternative market
potential functions
might perform better than (\ref{mdit}).

Table~\ref{tab:cfrmt} shows the $R^2$ and the corresponding
$\rho^2$ between observed and fitted values of instantaneous sales for
alternative models.
%
%t2 #&#
\begin{table}
\caption{Comparison among alternative model specifications for market
potential, $m(t)$}\label{tab:cfrmt}
\begin{tabular*}{\textwidth}{@{\extracolsep{\fill}}lcc@{}}
\hline
$\bolds{m(t)} $& $\bolds{R^2}$ & $\bolds{\rho^2}$ \\
\hline
(\ref{mdit})
& 0.999960 & 0.967295 \\
Constant market potential & 0.998766 & 0.877826 \\
(\ref{msenzaradice}) & 0.999930 & 0.964444 \\
(\ref{mgamma}) & 0.999931 & 0.964347 \\
\hline
\end{tabular*}
\end{table}
In detail, the formulations used were
%
%e5.3 #&#
\begin{equation}
\label{msenzaradice} m(t) = K \frac{1-e^{-(p_c+q_c)t}}{1+ ({q_c}/{p_c}) e^{-(p_c+q_c)t}}
\end{equation}
and
%
%e5.4 #&#
\begin{equation}
\label{mgamma} m(t) = K \cdot F(t)= K \cdot\int_0^t
\frac{1}{\Gamma(\alpha_1)}\alpha _0^{\alpha_1} t^{\alpha_1-1}
e^{- \alpha_0 t} \,{d}t,
\end{equation}
where $\Gamma(\alpha_1)=\int_0^\infty t^{\alpha_1-1} e^{-t} \,{d}t$,
and $\alpha_0$, $\alpha_1$, $t>0$.
The function in (\ref{msenzaradice}) represents a modification of the
proposed function (\ref{mdit}), while (\ref{mgamma}) describes
the evolution of the dynamic market potential as proportional to
$F(t)$, the cumulative distribution function of a Gamma random variable
(with mean equal to $\alpha_1/\alpha_0$ and variance equal to $\alpha
_1/\alpha_0^2)$.

The values presented in Table~\ref{tab:cfrmt} confirm that
a constant market potential assumption is not adequate to describe
these data. The structures
(\ref{msenzaradice}) and (\ref{mgamma}) perform slightly worse than the proposed
structure (\ref{mdit}).
However, the main difference is that a Gamma distribution or a
structure similar to
(\ref{msenzaradice}) only serves
the purpose of representing a flexible monotonic
function. Conversely, (\ref{mdit}) has been
proposed essentially because it represents the size of an informed
network spreading
information regarding the product category. Thus, this model structure
has a substantial
interpretative content. The proposed model is shown to perform best in
this application.
In light also of the results of the simulation study
proposed in Appendix 5 [\citeauthor{suppA} (\citeyear{suppA})], our
opinion is that
the CDMP model represents a useful contribution in the
field of competition diffusion of innovation models.

%s6 #&#
\section{Concluding remarks} \label{sec:discussion}

Diffusion of innovation methodologies
have faced and are facing
new challenges in parsimonious
model-building in terms of incorporating
the major effects that can modify the
evolutionary shapes of these methodologies over time.

This paper highlights the key features of the competition between
Amaryl and Solosa. These two drugs differ essentially
in the persuasion effects exerted by the two companies that launched
the drugs and in their acceptance through the community of physicians
spreading word-of-mouth about their efficacy.

The initial novelty of the active compound of these drugs in the market
suggested to
us that the existing models of competition must be enriched with the
introduction
of dynamic market potential. This extension rests on the concept that
awareness is a fundamental prerequisite for adoption.
We can imagine that, at the individual level, awareness and adoption
are two
sequential states that subjects (here, physicians)
may undergo. The first state, awareness, is latent. In addition, since
individual data
are unavailable in this case,
the description is aggregated (as a mean profile) and leads to equation
(\ref{mdit}).

Similarly, although in a very different
context, note that the \citet{guseo:2009} paper inspired the approach
followed by
\citet{SIM} to describe and predict the death toll due to pleural
mesothelioma contracted through exposure to asbestos fibers in a
residential area close
to a big plant. In that case, contamination (state 1)---that is,
contact with lethal asbestos fibers---was the latent prerequisite for
developing the disease (state 2).

Finally, we would
emphasize that our proposed model is useful specifically for analyzing
competition between two products. The tractability of the model,
in terms of the estimation
of the involved parameters, enables us to deal
with a higher number of competitors only if they have entered the
market simultaneously.
Diachronic competition, that is, among products launched at different times,
generally requires model structures
with multiple regimes (a change-point in the evolution of existing
products occurs
whenever a new competitor appears). In this case, for more than three products,
the parameter cardinality becomes too high to obtain reliable
estimates, unless each
regime is covered by an adequate observation period.

\section*{Acknowledgments}
The authors
are grateful to the Associate Editor and to the anonymous reviewers for
their helpful
suggestions that improved the quality of the manuscript.

\begin{supplement}[id=suppA]
\stitle{Supplementary materials}
\slink[doi]{10.1214/15-AOAS868SUPP} %[doi,text={...}] - jei reikia suskaldyti doi
\sdatatype{.pdf}
\sfilename{aoas868\_supp.pdf}
\sdescription{In Appendix 1 we provide details
regarding the closed-form solution of the proposed model.
In Appendix 2 we propose an alternative estimation method to deal with
monthly sales data instead of
cumulative observations. In Appendix 3 we discuss the construction of predictive
confidence bands. In Appendix 4 we present a SARMAX refinement for the
first-order model
fitting for short-term forecasting purposes. Finally, in Appendix 5 we
show the results of a simulation study to assess
the reliability of inferences.}
\end{supplement}

%\bibliographystyle{imsart-number}
%\bibliographystyle{imsart-nameyear}
%\bibliography{biblioAOAS868} %

\begin{thebibliography}{24}
% pybtex-1.44. Style name=ims, version=2.92, label_style=nameyear,
%sorting_style=complex, cfg=None, language=None.

%b1 ###
%b1 #&#
\bibitem[\protect\citeauthoryear{Abramson and Zanette}{1998}]{abramsoms:98}
%
\begin{barticle}[author]
\bauthor{\bsnm{Abramson},~\bfnm{G.}\binits{G.}} \AND
\bauthor{\bsnm{Zanette},~\bfnm{D.~H.}\binits{D.~H.}}
(\byear{1998}).
\btitle{Statistics of extinction and survival in {L}otka--{V}olterra systems}.
\bjournal{Phys. Rev. E (3)}
\bvolume{57}
\bpages{4572--4577}.
\end{barticle}
%
\iffalse\OrigBibText
%
\begin{barticle}[author]
\bauthor{\bsnm{Abramson},~\bfnm{G.}\binits{G.}} \AND
\bauthor{\bsnm{Zanette},~\bfnm{D.~H.}\binits{D.~H.}}
(\byear{1998}).
\btitle{Statistics of extinction and survival in
\uppercase{L}otka-\uppercase{V}olterra systems}.
\bjournal{Physical Review E}
\bvolume{57}
\bpages{4572--4577}.
\end{barticle}
%
\endOrigBibText\fi
\bptok{imsref}%
\endbibitem

%b2 ###
%b2 #&#
\bibitem[\protect\citeauthoryear{Bass}{1969}]{bass:69}
%
\begin{barticle}[author]
\bauthor{\bsnm{Bass},~\bfnm{F.~M.}\binits{F.~M.}}
(\byear{1969}).
\btitle{A new product growth model for consumer durables}.
\bjournal{Management Science}
\bvolume{15}
\bpages{215--227}.
\end{barticle}
%
\iffalse\OrigBibText
%
\begin{barticle}[author]
\bauthor{\bsnm{Bass},~\bfnm{F.~M.}\binits{F.~M.}}
(\byear{1969}).
\btitle{A new product growth model for consumer durables}.
\bjournal{Management Science}
\bvolume{15}
\bpages{215--227}.
\end{barticle}
%
\endOrigBibText\fi
\bptok{imsref}%
\endbibitem

%b3 ###
%b3 #&#
\bibitem[\protect\citeauthoryear{Bass, Krishnan and Jain}{1994}]{bass:94}
%
\begin{barticle}[author]
\bauthor{\bsnm{Bass},~\bfnm{F.}\binits{F.}},
\bauthor{\bsnm{Krishnan},~\bfnm{T.}\binits{T.}} \AND
\bauthor{\bsnm{Jain},~\bfnm{D.}\binits{D.}}
(\byear{1994}).
\btitle{Why the {B}ass model fits without decision variables}.
\bjournal{Marketing Science}
\bvolume{13}
\bpages{203--223}.
\end{barticle}
%
\iffalse\OrigBibText
%
\begin{barticle}[author]
\bauthor{\bsnm{Bass},~\bfnm{F.}\binits{F.}},
\bauthor{\bsnm{Krishnan},~\bfnm{T.}\binits{T.}} \AND
\bauthor{\bsnm{Jain},~\bfnm{D.}\binits{D.}}
(\byear{1994}).
\btitle{Why the \uppercase{B}ass model fits without decision variables}.
\bjournal{Marketing Science}
\bvolume{13}
\bpages{203--223}.
\end{barticle}
%
\endOrigBibText\fi
\bptok{imsref}%
\endbibitem

%b4 ###
%b4 #&#
\bibitem[\protect\citeauthoryear{Beauchamp and Cornell}{1966}]{beaucorn}
%
\begin{barticle}[mr]
\bauthor{\bsnm{Beauchamp},~\bfnm{John~J.}\binits{J.~J.}} \AND
\bauthor{\bsnm{Cornell},~\bfnm{Richard~G.}\binits{R.~G.}}
(\byear{1966}).
\btitle{Simultaneous nonlinear estimation}.
\bjournal{Technometrics}
\bvolume{8}
\bpages{319--326}.
\bid{issn={0040-1706}, mr={0205364}}
\end{barticle}
%
\iffalse\OrigBibText
%
\begin{barticle}[author]
\bauthor{\bsnm{Beauchamp},~\bfnm{J.~J.}\binits{J.~J.}} \AND
\bauthor{\bsnm{Cornell},~\bfnm{R.~G.}\binits{R.~G.}}
(\byear{1966}).
\btitle{Simultaneous nonlinear estimation}.
\bjournal{Technometrics}
\bvolume{8}
\bpages{319--326}.
\end{barticle}
%
\endOrigBibText\fi
\bptok{imsref}%
% NOT OUTPUTTED:
% fjournal = Technometrics. A Journal of Statistics for the Physical,
%Chemical and Engineering Sciences
\endbibitem

%b5 ###
%b5 #&#
\bibitem[\protect\citeauthoryear{Boswijk and Franses}{2005}]{Bosw}
%
\begin{barticle}[mr]
\bauthor{\bsnm{Boswijk},~\bfnm{H.~Peter}\binits{H.~P.}} \AND
\bauthor{\bsnm{Franses},~\bfnm{Philip~Hans}\binits{P.~H.}}
(\byear{2005}).
\btitle{On the econometrics of the {B}ass diffusion model}.
\bjournal{J.~Bus. Econom. Statist.}
\bvolume{23}
\bpages{255--268}.
\bid{doi={10.1198/073500104000000604}, issn={0735-0015}, mr={2159678}}
\end{barticle}
%
\iffalse\OrigBibText
%
\begin{barticle}[author]
\bauthor{\bsnm{Boswijk},~\bfnm{H.~P.}\binits{H.~P.}} \AND
\bauthor{\bsnm{Franses},~\bfnm{P.~H.}\binits{P.~H.}}
(\byear{2005}).
\btitle{On the Econometrics of the Bass Diffusion Model}.
\bjournal{Journal of Business and Economic Statistics}
\bvolume{23}
\bpages{255--268}.
\end{barticle}
%
\endOrigBibText\fi
\bptok{imsref}%
% NOT OUTPUTTED:
% number = 3
% doi = http://dx.doi.org/10.1198/073500104000000604
% fjournal = Journal of Business \& Economic Statistics
\endbibitem

%b6 ###
%b6 #&#
\bibitem[\protect\citeauthoryear{Centrone, Goia and Salinelli}{2007}]{Centrone}
%
\begin{barticle}[author]
\bauthor{\bsnm{Centrone},~\bfnm{Francesca}\binits{F.}},
\bauthor{\bsnm{Goia},~\bfnm{Aldo}\binits{A.}} \AND
\bauthor{\bsnm{Salinelli},~\bfnm{Ernesto}\binits{E.}}
(\byear{2007}).
\btitle{Demographic processes in a model of innovation diffusion with
dynamic market}.
\bjournal{Technological Forecasting and Social Change}
\bvolume{74}
\bpages{247--266}.
\end{barticle}
%
\iffalse\OrigBibText
%
\begin{barticle}[author]
\bauthor{\bsnm{Centrone},~\bfnm{Francesca}\binits{F.}},
\bauthor{\bsnm{Goia},~\bfnm{Aldo}\binits{A.}} \AND
\bauthor{\bsnm{Salinelli},~\bfnm{Ernesto}\binits{E.}}
(\byear{2007}).
\btitle{Demographic processes in a model of innovation diffusion with dynamic
market}.
\bjournal{Technological Forecasting and Social Change}
\bvolume{74}
\bpages{247--266}.
\end{barticle}
%
\endOrigBibText\fi
\bptok{imsref}%
\endbibitem

%b7 ###
%b7 #&#
\bibitem[\protect\citeauthoryear{Furlan and Mortarino}{2012}]{SIM}
%
\begin{barticle}[mr]
\bauthor{\bsnm{Furlan},~\bfnm{Claudia}\binits{C.}} \AND
\bauthor{\bsnm{Mortarino},~\bfnm{Cinzia}\binits{C.}}
(\byear{2012}).
\btitle{Pleural mesothelioma: Forecasts of the death toll in the area
of {C}asale {M}onferrato, {I}taly}.
\bjournal{Stat. Med.}
\bvolume{31}
\bpages{4114--4134}.
\bid{doi={10.1002/sim.5523}, issn={0277-6715}, mr={3041797}}
\end{barticle}
%
\iffalse\OrigBibText
%
\begin{barticle}[author]
\bauthor{\bsnm{Furlan},~\bfnm{C.}\binits{C.}} \AND
\bauthor{\bsnm{Mortarino},~\bfnm{C.}\binits{C.}}
(\byear{2012}).
\btitle{Pleural mesothelioma: forecasts of the death toll in the area
of Casale
Monferrato, Italy}.
\bjournal{Statistics in Medicine}
\bvolume{31}
\bpages{4114--4134}.
\end{barticle}
%
\endOrigBibText\fi
\bptok{imsref}%
% NOT OUTPUTTED:
% number = 29
% doi = http://dx.doi.org/10.1002/sim.5523
% fjournal = Statistics in Medicine
\endbibitem

%b8 ###
%b8 #&#
\bibitem[\protect\citeauthoryear{Guseo and Guidolin}{2009}]{guseo:2009}
%
\begin{barticle}[author]
\bauthor{\bsnm{Guseo},~\bfnm{R.}\binits{R.}} \AND
\bauthor{\bsnm{Guidolin},~\bfnm{M.}\binits{M.}}
(\byear{2009}).
\btitle{Modelling a dynamic market potential: A class of automata
networks for diffusion of innovations}.
\bjournal{Technological Forecasting and Social Change}
\bvolume{76}
\bpages{806--820}.
\end{barticle}
%
\iffalse\OrigBibText
%
\begin{barticle}[author]
\bauthor{\bsnm{Guseo},~\bfnm{R.}\binits{R.}} \AND
\bauthor{\bsnm{Guidolin},~\bfnm{M.}\binits{M.}}
(\byear{2009}).
\btitle{Modelling a dynamic market potential: a class of Automata
Networks for
diffusion of innovations}.
\bjournal{Technological Forecasting and Social Change}
\bvolume{76}
\bpages{806--820}.
\end{barticle}
%
\endOrigBibText\fi
\bptok{imsref}%
\endbibitem

%b9 ###
%b9 #&#
\bibitem[\protect\citeauthoryear{Guseo and Mortarino}{2012}]{EJOR}
%
\begin{barticle}[mr]
\bauthor{\bsnm{Guseo},~\bfnm{Renato}\binits{R.}} \AND
\bauthor{\bsnm{Mortarino},~\bfnm{Cinzia}\binits{C.}}
(\byear{2012}).
\btitle{Sequential market entries and competition modelling in
multi-innovation diffusions}.
\bjournal{European J. Oper. Res.}
\bvolume{216}
\bpages{658--667}.
\bid{doi={10.1016/j.ejor.2011.08.018}, issn={0377-2217}, mr={2845865}}
\end{barticle}
%
\iffalse\OrigBibText
%
\begin{barticle}[author]
\bauthor{\bsnm{Guseo},~\bfnm{R.}\binits{R.}} \AND
\bauthor{\bsnm{Mortarino},~\bfnm{C.}\binits{C.}}
(\byear{2012}).
\btitle{Sequential market entries and competition modelling in multi-innovation
diffusions.}
\bjournal{European Journal of Operational Research}
\bvolume{216}
\bpages{658--667}.
\end{barticle}
%
\endOrigBibText\fi
\bptok{imsref}%
% NOT OUTPUTTED:
% number = 3
% doi = http://dx.doi.org/10.1016/j.ejor.2011.08.018
% coden = EJORDT
% fjournal = European Journal of Operational Research
\endbibitem

%b10 ###
%b10 #&#
\bibitem[\protect\citeauthoryear{Guseo and Mortarino}{2014a}]{COST}
%
\begin{barticle}[mr]
\bauthor{\bsnm{Guseo},~\bfnm{Renato}\binits{R.}} \AND
\bauthor{\bsnm{Mortarino},~\bfnm{Cinzia}\binits{C.}}
(\byear{2014}a).
\btitle{Multivariate nonlinear least squares: Robustness and efficiency
of standard versus {B}eauchamp and {C}ornell methodologies}.
\bjournal{Comput. Statist.}
\bvolume{29}
\bpages{1609--1636}.
\bid{doi={10.1007/s00180-014-0509-y}, issn={0943-4062}, mr={3279009}}
\end{barticle}
%
\iffalse\OrigBibText
%
\begin{barticle}[author]
\bauthor{\bsnm{Guseo},~\bfnm{R.}\binits{R.}} \AND
\bauthor{\bsnm{Mortarino},~\bfnm{C.}\binits{C.}}
(\byear{2014}a).
\btitle{Multivariate nonlinear least squares: robustness and efficiency of
standard versus Beauchamp and Cornell methodologies}.
\bjournal{Computational Statistics}
\bvolume{29}
\bpages{1609--1636}.
\end{barticle}
%
\endOrigBibText\fi
\bptok{imsref}%
% NOT OUTPUTTED:
% number = 6
% doi = http://dx.doi.org/10.1007/s00180-014-0509-y
% fjournal = Computational Statistics
\endbibitem

%b11 ###
%b11 #&#
\bibitem[\protect\citeauthoryear{Guseo and Mortarino}{2014b}]{IMA}
%
\begin{barticle}[mr]
\bauthor{\bsnm{Guseo},~\bfnm{Renato}\binits{R.}} \AND
\bauthor{\bsnm{Mortarino},~\bfnm{Cinzia}\binits{C.}}
(\byear{2014}b).
\btitle{Within-brand and cross-brand word-of-mouth for sequential
multi-innovation diffusions}.
\bjournal{IMA J. Manag. Math.}
\bvolume{25}
\bpages{287--311}.
\bid{doi={10.1093/imaman/dpt008}, issn={1471-678X}, mr={3226506}}
\end{barticle}
%
\iffalse\OrigBibText
%
\begin{barticle}[author]
\bauthor{\bsnm{Guseo},~\bfnm{R.}\binits{R.}} \AND
\bauthor{\bsnm{Mortarino},~\bfnm{C.}\binits{C.}}
(\byear{2014}b).
\btitle{Within-brand and cross-brand word-of-mouth for sequential
multi-innovation diffusions}.
\bjournal{IMA Journal of Management Mathematics}
\bvolume{25}
\bpages{287--311}.
\end{barticle}
%
\endOrigBibText\fi
\bptok{imsref}%
% NOT OUTPUTTED:
% number = 3
% doi = http://dx.doi.org/10.1093/imaman/dpt008
% fjournal = IMA Journal of Management Mathematics
\endbibitem

%b12 #&#
\bibitem[\protect\citeauthoryear{Guseo and Mortarino}{2015}]{suppA}
%
\begin{bmisc}[author]
\bauthor{\bsnm{Guseo},~\bfnm{R.}\binits{R.}} \AND
\bauthor{\bsnm{Mortarino},~\bfnm{C.}\binits{C.}}
(\byear{2015}).
\bhowpublished{Supplement to ``Modeling competition between two
pharmaceutical drugs using innovation diffusion models.''
DOI:\doiurl{10.1214/15-AOAS868SUPP}}.
\bptok{imsref}%
\end{bmisc}
%
\endbibitem

%b13 ###
%b13 #&#
\bibitem[\protect\citeauthoryear{Handcock and Gile}{2010}]{hand:2010}
%
\begin{barticle}[mr]
\bauthor{\bsnm{Handcock},~\bfnm{Mark~S.}\binits{M.~S.}} \AND
\bauthor{\bsnm{Gile},~\bfnm{Krista~J.}\binits{K.~J.}}
(\byear{2010}).
\btitle{Modeling social networks from sampled data}.
\bjournal{Ann. Appl. Stat.}
\bvolume{4}
\bpages{5--25}.
\bid{doi={10.1214/08-AOAS221}, issn={1932-6157}, mr={2758082}}
\end{barticle}
%
\iffalse\OrigBibText
%
\begin{barticle}[author]
\bauthor{\bsnm{Handcock},~\bfnm{M.~S.}\binits{M.~S.}} \AND
\bauthor{\bsnm{Gile},~\bfnm{K.~J.}\binits{K.~J.}}
(\byear{2010}).
\btitle{Modeling social networks from sampled data}.
\bjournal{The Annals of Applied Statistics}
\bvolume{4}
\bpages{5--25}.
\end{barticle}
%
\endOrigBibText\fi
\bptok{imsref}%
% NOT OUTPUTTED:
% number = 1
% doi = http://dx.doi.org/10.1214/08-AOAS221
% fjournal = The Annals of Applied Statistics
\endbibitem

%b14 ###
%b14 #&#
\bibitem[\protect\citeauthoryear{Jha, Chaudhary and Gutpa}{2011}]{Jha:2011}
%
\begin{bmisc}[author]
\bauthor{\bsnm{Jha},~\bfnm{P.~C.}\binits{P.~C.}},
\bauthor{\bsnm{Chaudhary},~\bfnm{K.}\binits{K.}} \AND
\bauthor{\bsnm{Gutpa},~\bfnm{A.}\binits{A.}}
(\byear{2011}).
\bhowpublished{On the development of adoption of newer successive
technologies using stochastic differential equation.
In \textit{IEEE International Conference on Industrial Engineering and
Engineering Management}
1853--1858. IEEE, Singapore.}
\end{bmisc}
%
\iffalse\OrigBibText
%
\begin{barticle}[author]
\bauthor{\bsnm{Jha},~\bfnm{P.~C.}\binits{P.~C.}},
\bauthor{\bsnm{Chaudhary},~\bfnm{K.}\binits{K.}} \AND
\bauthor{\bsnm{Gutpa},~\bfnm{A}\binits{A.}}
(\byear{2011}).
\btitle{On the development of adoption of newer successive technologies using
stochastic differential equation}.
\bjournal{\uppercase{IEEE} International Conference on Industrial Engineering
and Engineering Management}
\bpages{1853--1858}.
\end{barticle}
%
\endOrigBibText\fi
\bptok{imsref}%
\endbibitem

%b15 ###
%b15 #&#
\bibitem[\protect\citeauthoryear{Kim, Bridges and Srivastava}{1999}]{Kim}
%
\begin{barticle}[author]
\bauthor{\bsnm{Kim},~\bfnm{Namwoon}\binits{N.}},
\bauthor{\bsnm{Bridges},~\bfnm{Eileen}\binits{E.}} \AND
\bauthor{\bsnm{Srivastava},~\bfnm{Rajendra~K.}\binits{R.~K.}}
(\byear{1999}).
\btitle{A simultaneous model for innovative product category sales
diffusion and competitive dynamics}.
\bjournal{International Journal of Research in Marketing}
\bvolume{16}
\bpages{95--111}.
\end{barticle}
%
\iffalse\OrigBibText
%
\begin{barticle}[author]
\bauthor{\bsnm{Kim},~\bfnm{Namwoon}\binits{N.}},
\bauthor{\bsnm{Bridges},~\bfnm{Eileen}\binits{E.}} \AND
\bauthor{\bsnm{Srivastava},~\bfnm{Rajendra~K.}\binits{R.~K.}}
(\byear{1999}).
\btitle{A simultaneous model for innovative product category sales diffusion
and competitive dynamics}.
\bjournal{International Journal of Research in Marketing}
\bvolume{16}
\bpages{95--111}.
\end{barticle}
%
\endOrigBibText\fi
\bptok{imsref}%
\endbibitem

%b16 ###
%b16 #&#
\bibitem[\protect\citeauthoryear{Krishnan, Bass and Kumar}{2000}]{krishnan:00}
%
\begin{barticle}[author]
\bauthor{\bsnm{Krishnan},~\bfnm{T.~V.}\binits{T.~V.}},
\bauthor{\bsnm{Bass},~\bfnm{F.~M.}\binits{F.~M.}} \AND
\bauthor{\bsnm{Kumar},~\bfnm{V.}\binits{V.}}
(\byear{2000}).
\btitle{Impact of a late entrant on the diffusion of a new product/service}.
\bjournal{Journal of Marketing Research}
\bvolume{XXXVII}
\bpages{269--278}.
\end{barticle}
%
\iffalse\OrigBibText
%
\begin{barticle}[author]
\bauthor{\bsnm{Krishnan},~\bfnm{T.~V.}\binits{T.~V.}},
\bauthor{\bsnm{Bass},~\bfnm{F.~M.}\binits{F.~M.}} \AND
\bauthor{\bsnm{Kumar},~\bfnm{V.}\binits{V.}}
(\byear{2000}).
\btitle{Impact of a late entrant on the diffusion of a new product/service}.
\bjournal{Journal of Marketing Research}
\bvolume{XXXVII}
\bpages{269--278}.
\end{barticle}
%
\endOrigBibText\fi
\bptok{imsref}%
\endbibitem

%b17 ###
%b17 #&#
\bibitem[\protect\citeauthoryear{Libai, Muller and Peres}{2009}]{libai:09}
%
\begin{barticle}[author]
\bauthor{\bsnm{Libai},~\bfnm{B.}\binits{B.}},
\bauthor{\bsnm{Muller},~\bfnm{E.}\binits{E.}} \AND
\bauthor{\bsnm{Peres},~\bfnm{R.}\binits{R.}}
(\byear{2009}).
\btitle{The role of within-brand and cross-brand communications in
competitive growth}.
\bjournal{Journal of Marketing}
\bvolume{73}
\bpages{19--34}.
\end{barticle}
%
\iffalse\OrigBibText
%
\begin{barticle}[author]
\bauthor{\bsnm{Libai},~\bfnm{B.}\binits{B.}},
\bauthor{\bsnm{Muller},~\bfnm{E.}\binits{E.}} \AND
\bauthor{\bsnm{Peres},~\bfnm{R.}\binits{R.}}
(\byear{2009}).
\btitle{The role of within-brand and cross-brand communications in competitive
growth}.
\bjournal{Journal of Marketing}
\bvolume{73}
\bpages{19--34}.
\end{barticle}
%
\endOrigBibText\fi
\bptok{imsref}%
\endbibitem

%b18 ###
%b18 #&#
\bibitem[\protect\citeauthoryear{Meade and Islam}{2006}]{meade}
%
\begin{barticle}[author]
\bauthor{\bsnm{Meade},~\bfnm{N.}\binits{N.}} \AND
\bauthor{\bsnm{Islam},~\bfnm{T.}\binits{T.}}
(\byear{2006}).
\btitle{Modelling and forecasting the diffusion of innovation---{A}
25-year review}.
\bjournal{International Journal of Forecasting}
\bvolume{22}
\bpages{519--545}.
\end{barticle}
%
\iffalse\OrigBibText
%
\begin{barticle}[author]
\bauthor{\bsnm{Meade},~\bfnm{N.}\binits{N.}} \AND
\bauthor{\bsnm{Islam},~\bfnm{T.}\binits{T.}}
(\byear{2006}).
\btitle{Modelling and forecasting the diffusion of innovation --
\uppercase{A}
25-year review}.
\bjournal{International Journal of Forecasting}
\bvolume{22}
\bpages{519--545}.
\end{barticle}
%
\endOrigBibText\fi
\bptok{imsref}%
\endbibitem

%b19 ###
%b19 #&#
\bibitem[\protect\citeauthoryear{Meyer and Ausubel}{1999}]{Meyer}
%
\begin{barticle}[author]
\bauthor{\bsnm{Meyer},~\bfnm{Perrin~S.}\binits{P.~S.}} \AND
\bauthor{\bsnm{Ausubel},~\bfnm{Jesse~H.}\binits{J.~H.}}
(\byear{1999}).
\btitle{Carrying capacity: A model with logistically varying limits}.
\bjournal{Technological Forecasting and Social Change}
\bvolume{61}
\bpages{209--214}.
\end{barticle}
%
\iffalse\OrigBibText
%
\begin{barticle}[author]
\bauthor{\bsnm{Meyer},~\bfnm{Perrin~S.}\binits{P.~S.}} \AND
\bauthor{\bsnm{Ausubel},~\bfnm{Jesse~H.}\binits{J.~H.}}
(\byear{1999}).
\btitle{Carrying capacity: a model with logistically varying limits}.
\bjournal{Technological Forecasting and Social Change}
\bvolume{61}
\bpages{209--214}.
\end{barticle}
%
\endOrigBibText\fi
\bptok{imsref}%
\endbibitem

%b20 ###
%b20 #&#
\bibitem[\protect\citeauthoryear{Peres, Muller and Mahajan}{2010}]{Peres2010}
%
\begin{barticle}[author]
\bauthor{\bsnm{Peres},~\bfnm{R.}\binits{R.}},
\bauthor{\bsnm{Muller},~\bfnm{E.}\binits{E.}} \AND
\bauthor{\bsnm{Mahajan},~\bfnm{V.}\binits{V.}}
(\byear{2010}).
\btitle{Innovation diffusion and new product growth models: A critical
review and research directions}.
\bjournal{International Journal of Research in Marketing}
\bvolume{27}
\bpages{91--106}.
\end{barticle}
%
\iffalse\OrigBibText
%
\begin{barticle}[author]
\bauthor{\bsnm{Peres},~\bfnm{R.}\binits{R.}},
\bauthor{\bsnm{Muller},~\bfnm{E.}\binits{E.}} \AND
\bauthor{\bsnm{Mahajan},~\bfnm{V.}\binits{V.}}
(\byear{2010}).
\btitle{Innovation diffusion and new product growth models: A critical review
and research directions}.
\bjournal{International Journal of Research in Marketing}
\bvolume{27}
\bpages{91--106}.
\end{barticle}
%
\endOrigBibText\fi
\bptok{imsref}%
\endbibitem

%b21 ###
%b21 #&#
\bibitem[\protect\citeauthoryear{Putsis}{1996}]{putsis}
%
\begin{barticle}[author]
\bauthor{\bsnm{Putsis},~\bfnm{W.~P.}\binits{W.~P.}}
(\byear{1996}).
\btitle{Temporal aggregation in diffusion models of first-time
purchase: Does choice of frequency matter?}
\bjournal{Technological Forecasting and Social Change}
\bvolume{51}
\bpages{265--279}.
\end{barticle}
%
\iffalse\OrigBibText
%
\begin{barticle}[author]
\bauthor{\bsnm{Putsis},~\bfnm{W.~P.}\binits{W.~P.}}
(\byear{1996}).
\btitle{Temporal aggregation in diffusion models of first-time
purchase: Does
choice of frequency matter?}
\bjournal{Technological Forecasting and Social Change}
\bvolume{51}
\bpages{265--279}.
\end{barticle}
%
\endOrigBibText\fi
\bptok{imsref}%
\endbibitem

%b22 ###
%b22 #&#
\bibitem[\protect\citeauthoryear{Savin and Terwiesch}{2005}]{savin}
%
\begin{barticle}[mr]
\bauthor{\bsnm{Savin},~\bfnm{Sergei}\binits{S.}} \AND
\bauthor{\bsnm{Terwiesch},~\bfnm{Christian}\binits{C.}}
(\byear{2005}).
\btitle{Optimal product launch times in a duopoly: Balancing life-cycle
revenues with product cost}.
\bjournal{Oper. Res.}
\bvolume{53}
\bpages{26--47}.
\bid{doi={10.1287/opre.1040.0157}, issn={0030-364X}, mr={2131098}}
\end{barticle}
%
\iffalse\OrigBibText
%
\begin{barticle}[author]
\bauthor{\bsnm{Savin},~\bfnm{S.}\binits{S.}} \AND
\bauthor{\bsnm{Terwiesch},~\bfnm{C.}\binits{C.}}
(\byear{2005}).
\btitle{Optimal product launch times in a duopoly: balancing life-cycle
revenues with product cost}.
\bjournal{Operations Research}
\bvolume{53}
\bpages{26--47}.
\end{barticle}
%
\endOrigBibText\fi
\bptok{imsref}%
% NOT OUTPUTTED:
% number = 1
% doi = http://dx.doi.org/10.1287/opre.1040.0157
% coden = OPREAI
% fjournal = Operations Research
\endbibitem

%b23 ###
%b23 #&#
\bibitem[\protect\citeauthoryear{Sharif and Ramanathan}{1981}]{Sharif}
%
\begin{barticle}[author]
\bauthor{\bsnm{Sharif},~\bfnm{M.~N.}\binits{M.~N.}} \AND
\bauthor{\bsnm{Ramanathan},~\bfnm{K.}\binits{K.}}
(\byear{1981}).
\btitle{Binomial innovation diffusion models with dynamic potential
adopter population}.
\bjournal{Technological Forecasting and Social Change}
\bvolume{20}
\bpages{63--87}.
\end{barticle}
%
\iffalse\OrigBibText
%
\begin{barticle}[author]
\bauthor{\bsnm{Sharif},~\bfnm{M.~N.}\binits{M.~N.}} \AND
\bauthor{\bsnm{Ramanathan},~\bfnm{K.}\binits{K.}}
(\byear{1981}).
\btitle{Binomial innovation diffusion models with dynamic potential adopter
population}.
\bjournal{Technological Forecasting and Social Change}
\bvolume{20}
\bpages{63--87}.
\end{barticle}
%
\endOrigBibText\fi
\bptok{imsref}%
\endbibitem

%b24 ###
%b24 #&#
\bibitem[\protect\citeauthoryear{Sydow and Schrey{\"{o}}gg}{2013}]{sydow}
%
\begin{bbook}[author]
\bauthor{\bsnm{Sydow},~\bfnm{J.}\binits{J.}} \AND
\bauthor{\bsnm{Schrey{\"{o}}gg},~\bfnm{G.}\binits{G.}}
(\byear{2013}).
\btitle{Self-Reinforcing Processes in and Among Organizations}.
\bpublisher{Palgrave MacMillan},
\blocation{New York}.
\end{bbook}
%
\iffalse\OrigBibText
%
\begin{bbook}[author]
\bauthor{\bsnm{Sydow},~\bfnm{J.}\binits{J.}} \AND
\bauthor{\bsnm{Schrey\"{o}gg},~\bfnm{G.}\binits{G.}}
(\byear{2013}).
\btitle{Self-Reinforcing Processes in and among Organizations}.
\bpublisher{Palgrave MacMillan, New York}.
\end{bbook}
%
\endOrigBibText\fi
\bptok{imsref}%
\endbibitem
\end{thebibliography}

% imsref loaded by akundreckaite, 2015-10-07 15:57:07
%

%\begin{appendix}
%\section{}
%\end{appendix}

%\begin{thebibliography}{99}
%\bibitem{r1}
%\bibitem{r1}
%\end{thebibliography}

\printaddresses
\end{document}